\documentclass[12pt,epsf]{article}
\usepackage{graphicx}
\usepackage{amssymb}

\begin{document}

\def\a{\alpha}
\def\b{\beta}
\def\c{\varepsilon}
\def\d{\delta}
\def\e{\epsilon}
\def\f{\phi}
\def\g{\gamma}
\def\h{\theta}
\def\k{\kappa}
\def\l{\lambda}
\def\m{\mu}
\def\n{\nu}
\def\p{\psi}
\def\q{\partial}
\def\r{\rho}
\def\s{\sigma}
\def\t{\tau}
\def\u{\upsilon}
\def\v{\varphi}
\def\w{\omega}
\def\x{\xi}
\def\y{\eta}
\def\z{\zeta}
\def\D{\Delta}
\def\G{\Gamma}
\def\H{\Theta}
\def\L{\Lambda}
\def\F{\Phi}
\def\P{\Psi}
\def\S{\Sigma}
\def\BR{{\rm Br}}
\def\o{\over}
\def\beq{\begin{eqnarray}}
\def\eeq{\end{eqnarray}}
\newcommand{\nn}{\nonumber \\}
\newcommand{\gsim}{ \mathop{}_{\textstyle \sim}^{\textstyle >} }
\newcommand{\lsim}{ \mathop{}_{\textstyle \sim}^{\textstyle <} }
\newcommand{\vev}[1]{ \left\langle {#1} \right\rangle }
\newcommand{\bra}[1]{ \langle {#1} | }
\newcommand{\ket}[1]{ | {#1} \rangle }
\newcommand{\EV}{ {\rm eV} }
\newcommand{\KEV}{ {\rm keV} }
\newcommand{\MEV}{ {\rm MeV} }
\newcommand{\GEV}{ {\rm GeV} }
\newcommand{\TEV}{ {\rm TeV} }
\def\diag{\mathop{\rm diag}\nolimits}
\def\Spin{\mathop{\rm Spin}}
\def\SO{\mathop{\rm SO}}
\def\O{\mathop{\rm O}}
\def\SU{\mathop{\rm SU}}
\def\U{\mathop{\rm U}}
\def\Sp{\mathop{\rm Sp}}
\def\SL{\mathop{\rm SL}}
\def\tr{\mathop{\rm tr}}

\newcommand{\bear}{\begin{array}}  
\newcommand {\eear}{\end{array}}
\newcommand{\la}{\left\langle}  
\newcommand{\ra}{\right\rangle}
\newcommand{\non}{\nonumber}  
\newcommand{\ds}{\displaystyle}
\newcommand{\red}{\textcolor{red}}
\def\ubl{U(1)$_{\rm B-L}$}
\def\REF#1{(\ref{#1})}
\def\lrf#1#2{ \left(\frac{#1}{#2}\right)}
\def\lrfp#1#2#3{ \left(\frac{#1}{#2} \right)^{#3}}
\def\OG#1{ {\cal O}(#1){\rm\,GeV}}

\def\TODO#1{ {\bf ($\clubsuit$ #1 $\clubsuit$)} }

\newcommand{\rem}[1]{{\bf #1}}


\baselineskip 0.7cm

\begin{titlepage}

\begin{flushright}
UT-12-37 \\
IPMU12-0208
\end{flushright}

\vskip 1.35cm
\begin{center} 
{\large \bf Enhanced Higgs Mass in a Gaugino Mediation Model 
\\without the Polonyi Problem}
\vskip 1.2cm

{Takeo Moroi$^{(a)(b)}$, Tsutomu T. Yanagida$^{(b)}$ and  Norimi Yokozaki$^{(b)}$}

\vskip 0.4cm

{\it
$^{(a)}${Department of Physics, University of Tokyo,
   Tokyo 113-0033, Japan}\\
$^{(b)}${Kavli IPMU, University of Tokyo, Kashiwa, 277-8583, Japan}
}

\vskip 1.5cm

\abstract{ We consider a SUSY breaking scenario without the Polonyi
  problem. To solve the problem, the enhanced couplings of the Polonyi
  field to an inflaton, gauge kinetic functions and itself are
  assumed. As a result, a gaugino mediated SUSY breaking occurs.  In
  this scenario, the Higgs boson mass becomes consistent with the
  recently observed value of the Higgs-like boson (i.e., $m_h\simeq
  125\ {\rm GeV}$) for the gluino mass about 4\,TeV, which is,
  however, out of the reach of the LHC experiment.  We show that the
  trilinear coupling of the scalar top is automatically enhanced by
  the presence of the extra matters.  With such extra matters, the
  Higgs mass as large as $125\ {\rm GeV}$ can be realized with the
  gluino mass of $1-2$ TeV which is within the reach of the LHC
  experiment.  In our scenario, the gravitino is the lightest SUSY
  particle and the candidate for dark matter, and the Wino, Bino, and
  sleptons are in a range from 200 GeV to 700 GeV. }
\end{center}
\end{titlepage}

\setcounter{page}{2}

\section{Introduction}
The presence of so called Polonyi field $Z$ is an inevitable
ingredient in the gravity mediation when the gravitino mass is
$\lesssim\mathcal{O}(1)$\,TeV, otherwise we have vanishing gaugino
masses at the tree level and the contributions from the anomaly
mediation are too small~\cite{AMSB2, AMSB}.\footnote{
  In scenarios with $\mathcal{O}(10-100)$ TeV gravitino, the Polonyi
  field is not required since the gaugino masses of $\mathcal{O}(1)$
  TeV are induced at one-loop level. Without assuming a particular
  form of the Kahler potential, "Pure Gravity Mediation"
  scenario~\cite{PureGM, Pure2} was proposed where the scalar masses
  are $\mathcal{O}(10-100)$\,TeV and the gaugino masses are
  $\mathcal{O}(1)$\,TeV. (A similar model was proposed in
  Ref.~\cite{spreadSUSY}.) The scenario can naturally explain the
  Higgs mass of around 125 GeV with heavy stops~\cite{Higgs1}.  }
However, this Polonyi field $Z$ causes serious cosmological
problems. In particular, its decay in the early universe produces too
much entropy, resulting in a huge dilution of the primordial
baryon-number asymmetry. Furthermore, its decay occurs during/after
the Big-Bang Nucleosynthesis (BBN) and destroys light elements
produced by the BBN \cite{Polonyi}. From a cosmological view point,
the Polonyi problem severely restricts the supersymmetry (SUSY)
breaking scenarios.

In a series of recent works, it has been pointed out \cite{NTY1, NTY2}
that the serious Polonyi problem can be solved if the Polonyi field
has enhanced couplings to inflaton. This observation is based on the
adiabatic evolution of the Polonyi field following the inflaton
potential, originally suggested by Linde~\cite{Linde:1996cx}. The
integration of inflaton field induces also enhanced self couplings of
the Polonyi field; the Polonyi mass becomes much heavier than the
gravitino mass.  In addition, enhanced couplings of the Polonyi field
$Z$ to the gauge kinetic function are preferred in order to solve the
Polonyi problem. This is because the enhanced couplings make the decay
of $Z$ faster and the cosmological constraint becomes
weaker~\cite{KKM}. Consequently, relatively high reheating temperature
$T_R\sim 10^{6}$ GeV is allowed~\cite{NTY2}. With this reheating
temperature, non-thermal leptogenesis works for baryogenesis
\cite{nonthermal}.

Such a setup suggests a gaugino mediated SUSY breaking scenario
\cite{gaugino_med}.\footnote{If we assume a separation between the
  Polonyi field $Z$ and quark, lepton and Higgs multiplets in the
  conformal frame, we have vanishing soft masses for
  sfermions~\cite{gaugino_med}. Based on this observation, the gaugino
  mediation scenario was first proposed in \cite{gaugino_med} (see
  also ~\cite{MSYY}).} (See Refs.~\cite{gaugino2} for a scenario in
the framework of extra dimension.)  Here, we consider the case where
the interactions between the gauge multiplets and the Polonyi field
are enhanced while those between chiral multiplets in the minimal
supersymmetic standard model (MSSM) and the Polonyi field are not.  We
study its phenomenological consequences paying particular attention to
the Higgs mass.  As well as the cosmological advantages mentioned
above, such a scenario is also favored because it can solve the
serious SUSY FCNC problem~\cite{gaugino_med}.  As we discuss below, in our setup, the
gravitino mass $m_{3/2}$ is much smaller than the gaugino and the
Polonyi masses because of the enhanced couplings.\footnote
{A similar setup was considered in \cite{Buchmuller:2005rt}.}
The sfermions masses and the scalar trilinear couplings (so-called
$A$-terms) are also of the order of the gravitino mass at the
tree-level,\footnote
{If we assume the sequestered form of the Kahler potential, sfermion
  masses and $A$-terms vanish at the tree level~\cite{gaugino_med, AMSB2}.}
and dominantly arise from the gaugino masses through the
renormalization group evolution. Therefore, the SUSY CP and flavor
problems are relaxed.
 
In this letter, motivated by the recent discovery of the Higgs-like
boson at the ATLAS and CMS experiments~\cite{ATLAS_CMS}, we
investigate the Higgs boson mass in the gaugino mediated SUSY breaking
scenario.  If the particle content of the MSSM is assumed up to the
cut-off scale (which will be taken to be at the GUT scale), the Higgs
mass of around 125 GeV is realized only in the region out of the reach of
the LHC experiment \cite{Brummer:2012ns}. However, we point out that
the existence of the extra matters can drastically enhance the
trilinear coupling of the stop, even if there are no direct couplings
to the MSSM matter fields.  As a result, the Higgs mass can be as
large as $125$ GeV with large $A$-term~\cite{Higgs2} in a region
within the reach of the LHC experiment; gluino mass is around
1$-$2\,TeV. The existence of the extra matters at the scale much lower
than the Planck scale is expected in many models, for example, like
$E_6$ grand unified theory, axion models, and so on.

This paper is organized as follows. In section 2, we explain the setup
of our ``gaugino dominated SUSY breaking" scenario and show that the
Higgs mass of 125 GeV can be consistent only with the gluino mass
heavier than about 4 TeV with the particle content of the
MSSM~\cite{Brummer:2012ns}. In section 3, we show that the trilinear
coupling of the stop is enhanced if the extra matters exist; the Higgs
mass can be explained with the gluino mass of $1-2$ TeV. The final
section is devoted to conclusion and discussion.

%
%
%
%
\section{Gaugino mediation without Polonyi problem}
Let us discuss the setup of our gaugino mediation scenario. We assume
that the Polonyi field $Z$ strongly couples to the inflaton field and
the gauge kinetic functions, while the couplings of $Z$ to the other
fields (matters and Higgs) are suppressed. The relevant part of the
Kahler potential is given by
\begin{eqnarray}
K = - c_I^2 \frac{|Z|^2 |I|^2}{M_{P}^2} - \frac{c_Z^2 |Z|^4}{4 M_{P}^2},
\end{eqnarray}
where $Z$ and $I$ are the Polonyi field and the inflaton field,
respectively, and $M_P \simeq 2.4 \times 10^{18}\, {\rm GeV}$ is the reduced Plank scale. The coefficient $c_I$ is required to be as large as
$\sim100$ so that the Polonyi abundance is sufficiently suppressed by
the adiabatic suppression mechanism~\cite{Linde:1996cx, NTY1,
  NTY2}. The second term arises by the radiative corrections from the
inflaton loops and $c_Z$ is also expected to be $\sim100$. As a
result, the Polonyi mass is enhanced compared to the gravitino mass:
\begin{eqnarray}
{m_Z} = c_Z \frac{F_Z}{M_{P}} \simeq c_Z \sqrt{3} m_{3/2},
\end{eqnarray}
where $m_{3/2}$ is the gravitino mass. Here we assume that the SUSY is dominantly broken by the $Z$ field so that $m_{3/2} \simeq F_Z/\sqrt{3} M_P$. The couplings of $Z$ to the gauge kinetic functions are  also assumed to be enhanced:
\begin{eqnarray}
\mathcal{L} \ni -c_g \int d^2\theta \frac{ Z W_{\alpha} W^{\alpha}}{M_{P}} + h.c. \, , \label{eq:gaugino}
\end{eqnarray}
where $c_g \sim 100$. (Here we take the basis in which the gauge kinetic function is canonically normalized.)
With Eq.\ (\ref{eq:gaugino}), the gaugino mass is given by
\begin{eqnarray}
M_{\lambda} = 2 c_g \frac{F_Z}{M_P}  \simeq 2 \sqrt{3} c_g m_{3/2}. 
\end{eqnarray}
Since we consider the scenario that $c_g \sim c_Z \sim 100$, the
gravitino is the LSP and candidate for a dark matter. The Polonyi
field decays into SM gauge bosons through the operator
(\ref{eq:gaugino}), and its decay width is given by
\begin{eqnarray}
\Gamma(Z  \to {\rm 2\ gauge\,\,  bosons}) \simeq c_g^2 \frac{3 m_Z^3 }{2 \pi M_P^2} \simeq 1.2\,  {\rm sec}^{-1} \left(\frac{c_g}{100}\right)^2 \left(\frac{m_Z}{10^3\,{\rm GeV}}\right)^3.
\end{eqnarray}
With the suppression of the Polonyi abundance and the relatively short
lifetime (less than 1 second), the BBN constraint can be avoided
relatively easily~\cite{KKM} and the reheating temperature $T_R \simeq
10^6$\,GeV is allowed for $m_{3/2} \gtrsim 30$\,GeV and $c_I \simeq c_Z \simeq
c_g \simeq 100$~\cite{NTY2}. With such a relatively high reheating
temperature, enough baryon asymmetry can be generated by the
non-thermal leptogenesis~\cite{nonthermal}.

\begin{figure}[t]
\begin{center}
\includegraphics[width=7.9cm]{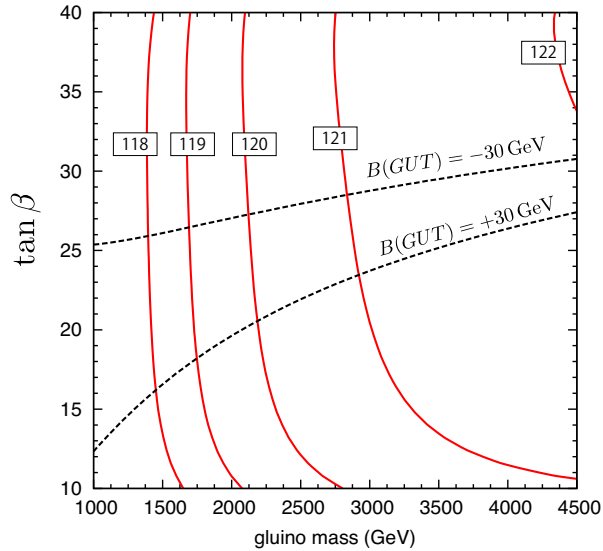}
\caption{The contours of the Higgs boson mass and the Higgs $B$-parameter at the GUT scale. The NLSP is the lightest stau. Here, $\alpha_S(m_Z)=0.1184$ and $m_t=173.2$ GeV.}
\label{fig:mssm}
\end{center}
\end{figure}

In our setup, the gravitino mass as well as the scalar masses are
suppressed compared to the gaugino masses at the cut-off scale. Then
the scalar masses dominantly arise from the renormalization group
effect between the cut-off scale and the SUSY scale (which is the mass
scale of the MSSM SUSY particles) of $\mathcal{O}(1)$\,TeV.  We
calculate the low-energy SUSY parameters by numerically solving
two-loop renormalization group equations (RGEs). Here, we use {\tt
  SuSpect}~\cite{suspect} to evaluate the spectrum of the SUSY
particles. The boundary condition is taken such that all the scalar
masses and trilinear coupling constants vanish and the gaugino masses
are universal at the cut-off scale (which is taken to be the GUT
scale). Then, we calculate the lightest Higgs boson mass using {\tt
  FeynHiggs}~\cite{FeynHiggs}.  In fig.~\ref{fig:mssm}, the contours
of the constant Higgs mass are shown.  In the same figure, we also
plot the contours of constant $B$-parameter at the GUT scale, where
the $B$-parameter is defined as
\begin{eqnarray}
  V \ni B \mu H_u H_d + h.c.,
\end{eqnarray}
with $H_u$ ($H_d$) being the up-type (down-type) Higgs and $\mu$ being
the Higgsino mass parameter in the superpotential.  (Here, $\mu$ is
assumed to be a free parameter, and is added by hand.)  The GUT-scale
value of the $B$-parameter (which is denoted as $B_{\rm GUT}$) is
expected to be of the same order of the gravitino mass.  As we have
mentioned, $m_{3/2} \gtrsim 30$\,GeV is required for non-thermal
leptogenesis with avoiding the BBN constraints.  So, we take $B_{\rm
  GUT}=\pm 30$\,GeV as representative values.

Adapting the uncertainty in the Higgs mass calculation of $2-3$
GeV~\cite{delta_mh, three_loop},
Higgs mass as large as $m_h \simeq 125$ GeV may be realized if the
gluino mass is heavier than about $4$\,TeV~\cite{Brummer:2012ns}.
Unfortunately, such a heavy gluino (and squarks) is out of the reach
of the LHC experiments. 

\section{Enhanced $A$-term from extra matters}

So far, we have seen that, if we adopt the particle content of the
MSSM, the Higgs mass of $m_h\simeq 125\ {\rm GeV}$ is hardly realized
in gaugino mediation model with gluino and squarks which are within
the reach of the LHC experiment.  Now, we show that such a conclusion
is altered if there exist extra vector-like multiplets at around
$1-10$ TeV.  Many models such as $E_6$ grand unified theory and axion
models predict the existence of the extra matters at the scale much
lower than the Planck scale.

The existence of extra matters may enhance the Higgs mass via two
effects.  First, if the extra matters have sizable Yukawa interaction
with the Higgs fields, the radiative correction below the SUSY scale
may significantly enhance the lightest Higgs mass \cite{vector1}.
This is the case with ${\bf 10}+{\bf \overline{10}}$ extra matters,
for example.  Second, the presence of the vector-like multiplets
changes the beta-functions of the gauge couplings and the
gauginos. Consequently, the trilinear coupling of the stop becomes
larger than the case without extra matters, resulting in the
enhancement of the Higgs mass.  The first effect has been intensively
studied in recent works \cite{vector2}, so we concentrate on the
second one, assuming that the extra matters have no Yukawa interaction with the Higgs fields.

First, we show how $A$-parameters, a squark mass and the ratio of the
$A$-parameter to the squark mass $A_{q}/m_{\tilde{q}}$ are enhanced
with extra matters.  For this purpose, we use one-loop RGEs (although
our numerical calculations are performed at the two-loop level).  With
the presence of the extra matters, the beta-functions of the gauge
coupling constants and gaugino masses are given by
\begin{eqnarray}
  \frac{d g_a^2}{d \ln Q} &=& (b_{a}+N_5)\frac{g_a^4}{8\pi^2},
  \\
  \frac{d M_a}{d \ln Q} &=& (b_{a}+N_5)\frac{g_a^2}{8\pi^2} M_a,
\end{eqnarray}
where $g_a\, (a=1,2,3)$ are the gauge couplings for $U(1)_Y$,
$SU(2)_L$ and $SU(3)_C$, respectively, and $b_a=(33/5, 1, -3)$ are the
coefficients of the beta-functions with the MSSM matter content. The
number of the extra vector-like multiplets in units of fundamental and
anti-fundamental representation of $SU(5)$ GUT gauge group, ${\bf 5} +
\overline{\bf 5}$, is denoted as $N_5$. For $N_5 \gtrsim 3$, $g_3$ and
$M_3$ at the SUSY scale are smaller than those at the GUT
scale.\footnote
{Although the one-loop beta-function of $g_3$ vanishes for $N_5=3$,
  inclusion of the higher order corrections leads to the positive
  beta-function. }
The change of these beta-functions dramatically alters the squark
masses and trilinear couplings at the SUSY scale. In particular, the
changes of the stop masses and stop trilinear coupling lead to the
important consequences in the Higgs boson mass and the SUSY search at
the LHC.

Neglecting Yukawa couplings, the RGE of an $A$-parameter of a squark
can be written as
\begin{eqnarray}
  \frac{d A_q}{d \ln Q} &=& 
  \frac{1}{16\pi^2} \left(c_a g_a ^2 M_a \right) 
  \nonumber \\
  &=& 
  \frac{1}{16\pi^2} \left[c_a \left( \frac{8\pi^2}{b_a + N_5} \right) 
    \frac{d g_a^2}{d \ln Q} \left(\frac{M_a}{g_a^2}  \right) \right].
  \label{Adot}
\end{eqnarray}
Notice that $M_a/g_a^2$ is an RGE invariant quantity, i.e.,
constant. The coefficient $c_3=32/3$ is common to all $A_q$.  (Here,
we neglect the effects of Yukawa coupling constants, which do not
change the following discussion qualitatively.  Our numerical
calculation will be performed with the effects of Yukawa coupling
constants.)  Eq.\ (\ref{Adot}) can be solved as
\begin{eqnarray}
A(Q) &=& -\frac{c_a}{2 (b_a + N_5)} \left[g_a^2(M_{\rm GUT})-g_a^2(Q)\right] \left(\frac{M_a}{g_a^2}\right)  \ \ \nonumber \\
&=& -\frac{c_a}{2} \left( \frac{g_a^4(Q)}{8\pi^2} \ln \frac{M_{\rm GUT}}{Q} \right) \left(1 -  \frac{b_a + N_5}{8\pi^2} g_a^2 (Q) \ln \frac{M_{\rm GUT}}{Q}\right)^{-1} \left(\frac{M_a}{g_a^2}\right).
\end{eqnarray}
Taking the renormalization scale $Q$ as the mass of the extra matter,
$M_{N_5}$, i.e., the decoupling scale, larger $N_5$ results in a
larger $A_q(M_{N_5})$ for the fixed value of $(M_a/g_a^2)$, since
$g_a(M_{N5})$ does not depend on $N_5$; the trilinear couplings
including $A_t$ are enhanced for the fixed gluino mass.

Similarly, the RGE of a squark mass can be written as
\begin{eqnarray}
\frac{d m_{\tilde{q}}^2}{d \ln Q} &=& \frac{1}{16\pi^2} \left(-d_a g_a ^2 M_a^2 \right) \nonumber \\
&=& \frac{1}{16\pi^2} \left[-d_a \left( \frac{4\pi^2}{b_a + N_5} \right) \frac{d g_a^4}{d \ln Q} \left(\frac{M_a}{g_a^2}  \right)^2 \right], \label{eq:msq_RGE}
\end{eqnarray}
where $d_3=32/3$ is common to all squark mass.  By solving
Eq.\,(\ref{eq:msq_RGE}), we obtain
\begin{eqnarray}
m_{\tilde{q}}^2(Q) &=& \frac{1}{4}\left(\frac{d_a}{b_a + N_5}\right) \left[ g_a^4(M_{\rm GUT})-g_a^4(Q) \right] \left(\frac{M_a}{g_a^2}  \right)^2 \nonumber \\
&=& \frac{d_a g_a^4(Q)}{4} \left( \frac{g_a^2(Q)}{8\pi^2} \ln \frac{M_{\rm GUT}}{Q} \right) \Biggl[ 
\left(1-\frac{b_a + N_5}{8\pi^2} g_a^2(Q) \ln \frac{M_{\rm GUT}}{Q}\right)^{-2} \nonumber \\
&&+ \left(1-\frac{b_a + N_5}{8\pi^2} g_a^2(Q) \ln \frac{M_{\rm GUT}}{Q} \right)^{-1} \Biggr] \left(\frac{M_a}{g_a^2}  \right)^2. 
\end{eqnarray}
Again, $m_{\tilde{q}}^2(M_{N_5})$ becomes larger as $N_5$ increases
for the fixed value of $M_a/g_a^2$. Thus by adding extra matters, the
stop mass $m_{\tilde{t}}$ is expected to be larger with the fixed
value of the gluino mass at the SUSY scale.  The ratio
$A_q(M_{N_5})/m_{\tilde{q}}(M_{N_5})$ is also enhanced. Neglecting the
contributions from $g_1$ and $g_2$, the ratio
$(A_q(Q)/m_{\tilde{q}}(Q))^2$ becomes
\begin{eqnarray}
\left( \frac{A(Q)}{m_{\tilde{q}}(Q)} \right)^2 &=& \frac{c_3^2}{d_3 (b_3 + N_5)} \left(1-g_3^2(Q)/g_3^2(M_{\rm GUT})\right) \left(1+g_3^2(Q)/g_3^2 (M_{\rm GUT}) \right)^{-1} \nonumber \\
&=& \left(\frac{c_3^2}{d_3}\right) \left( \frac{g^2(Q)}{8\pi^2} \ln \frac{M_{\rm GUT}}{Q} \right) \left( 2- \frac{b_3 + N_5}{8\pi^2} g^2(Q) \ln \frac{M_{\rm GUT}}{Q} \right)^{-1} .
\end{eqnarray}
Taking $Q=M_{N_5}$, we obtain the enhanced ratio
$[A_q(M_{N_5})/m_{\tilde{q}}(M_{N_5})]^2$ for larger
$N_5$. Consequently, the Higgs boson mass is enhanced as the number of
the extra matters increases, because of the larger ratio of
$A_t/m_{\tilde{t}}$ and the larger $m_{\tilde{t}}$.

\begin{figure}[t]
\begin{center}
\includegraphics[width=7.9cm]{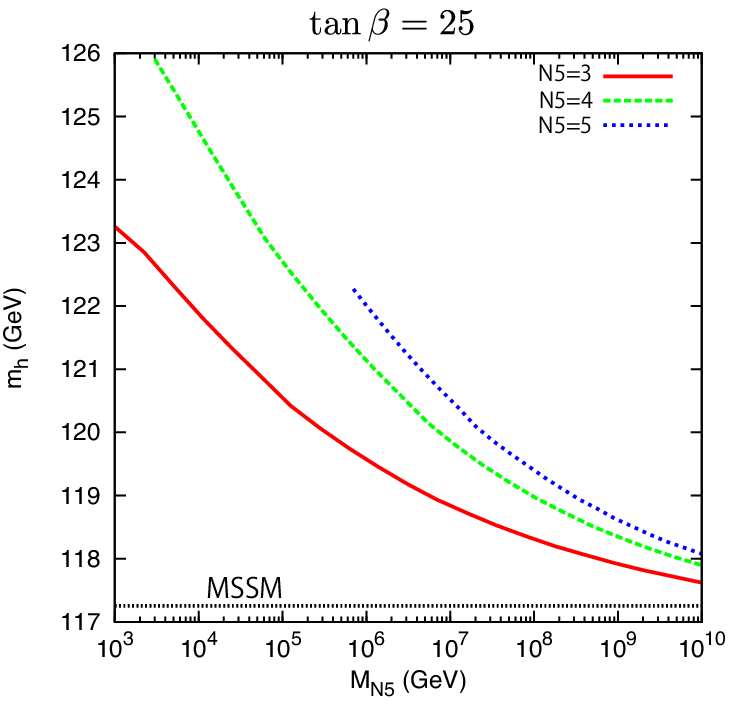}
\includegraphics[width=7.9cm]{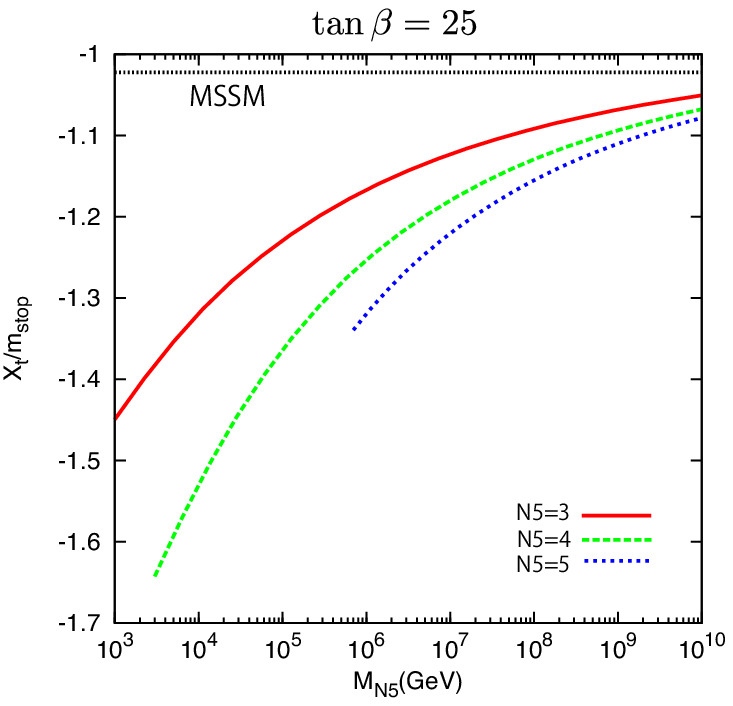}
\caption{The Higgs boson mass and the normalized trilinear coupling of
  the stop as a function of the decoupling scale of the extra
  matter. The gluino mass is fixed to be $m_{\tilde{g}}=1.2\, {\rm
    TeV}$. Here, $\tan\beta=25$. }
\label{fig:mn5_higgs}
\end{center}
\end{figure}

The results of the numerical calculations are shown in
fig.~\ref{fig:mn5_higgs}. The gluino mass is fixed to be 1.2 TeV and
$\tan\beta=25$ in both left and right panels. In the left panel, the
Higgs mass as a function of $M_{N_5}$ is shown.  Three curves
correspond to $N_5=3,4,5$ from bottom to top. For comparison, the
dashed line, which is evaluated in the MSSM is also drawn. Remarkably, the
Higgs mass reaches 125 GeV with $N_5=4$ and $M_{N_5} \simeq 3$ TeV. In
the right panel, we also show the normalized trilinear coupling
$X_t/m_{\tilde{t}}$, where $m_{\tilde{t}}\equiv (m_{\tilde{t}_1} +
m_{\tilde{t}_2})/2$ and $X_t=A_t-\mu \cot\beta$, with
$m_{\tilde{t}_1}$ and $m_{\tilde{t}_2}$ being the lighter and heavier
stop masses. Three curves correspond to $N_5=3,4,5$ from top to
bottom. The enhancement of $X_t/m_{\tilde{t}}$ can be seen. The number
of the vector-like multiplets $N_5=3$ is motivated by the $E_6$ grand
unified theory, and $N_5=4$ can be regarded as a complete vector-like
family.

\begin{figure}[t]
\begin{center}
\includegraphics[width=7.9cm]{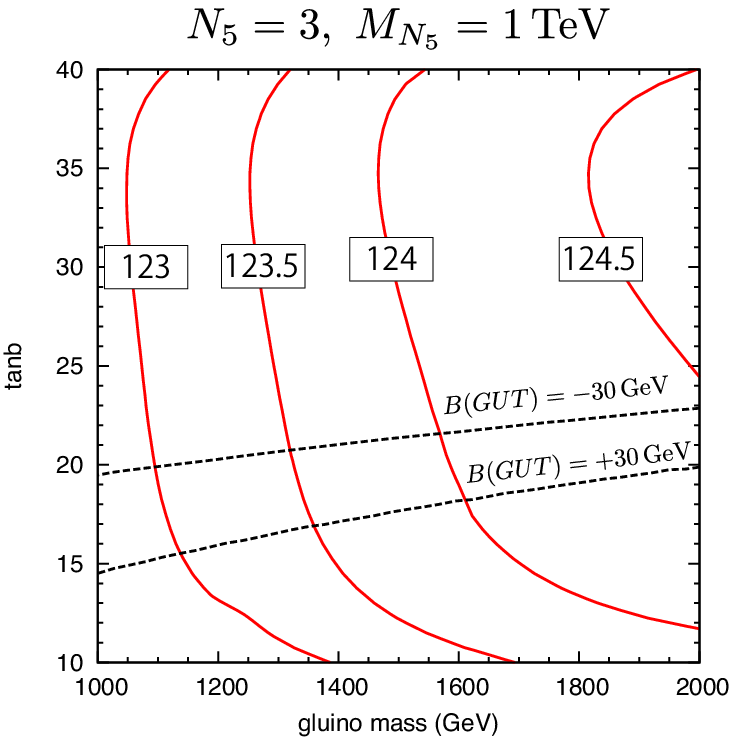}
\includegraphics[width=7.9cm]{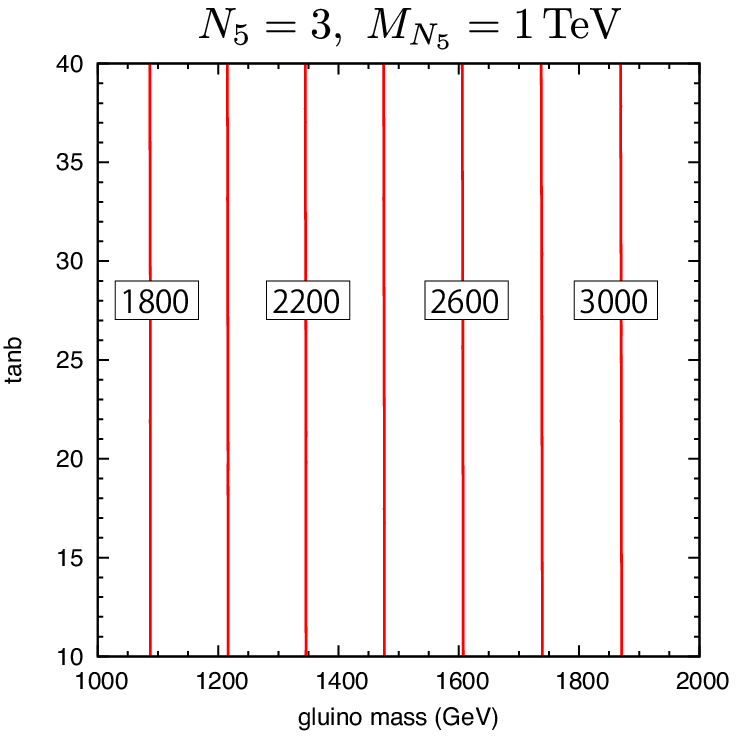}
\caption{Contours of the Higgs mass (left) and the squark mass (right)
  on $m_{\tilde{g}} - \tan\beta$ plane in the unit of GeV.  The three
  pairs of the extra matters exist at $1$ TeV. The vanishing A-terms
  and scalar masses are taken at the GUT scale. }
\label{fig:nf3}
\end{center}
\end{figure}

The contours of the constant Higgs mass for $N_5=3$ and $M_{N_5}=1$
TeV are shown in fig.~{\ref{fig:nf3}} (left panel). The regions near
the lines $B_{\rm GUT} = \pm 30$ GeV are consistent with the boundary
condition of the gaugino mediated SUSY breaking scenario. The Higgs
mass is calculated to be $123-124$ GeV for the gluino mass of
$1.1-1.6$ TeV. Correspondingly, a squark mass is $1.8-2.6$ TeV (right
panel). Notice that these gluino and squarks may be observed at the LHC.
%
%
In this region, the mass of the (right-handed) slepton is
about $470-670$ GeV and the Bino (Wino) mass is $210-310$ GeV
($330-490$ GeV), provided the universal gaugino masses at the GUT
scale. Such relatively light non-colored SUSY particles may be seen at future $e^+ e^-$ linear collider experiments.

\begin{figure}[t]
\begin{center}
\includegraphics[width=7.9cm]{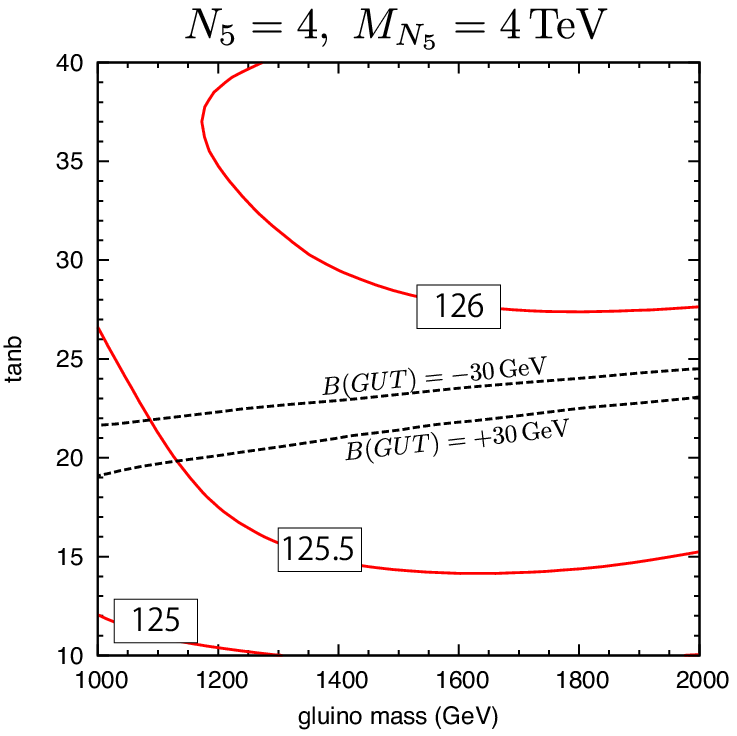}
\includegraphics[width=7.9cm]{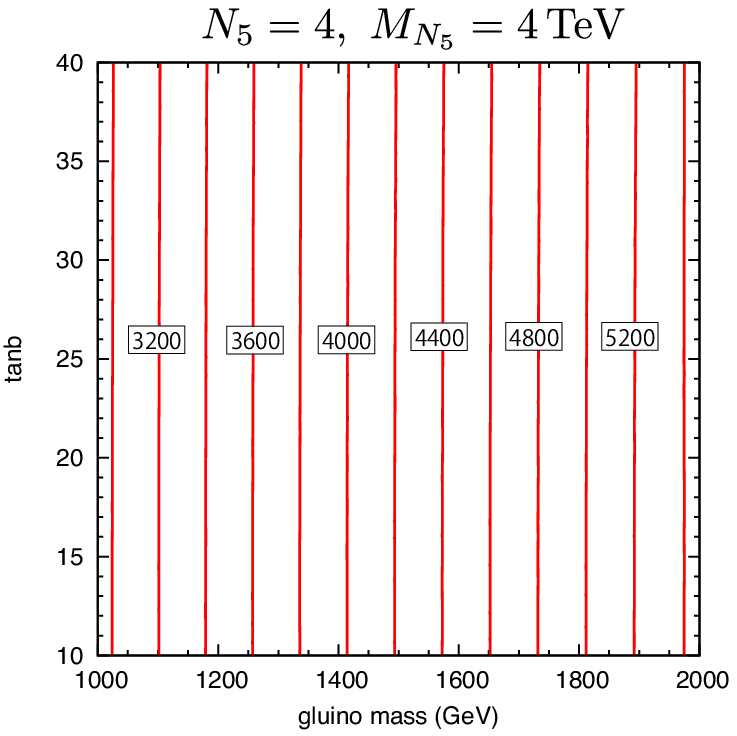}
\caption{Contours of the Higgs mass (left) and the squark mass (right)
  on $m_{\tilde{g}} - \tan\beta$ plane in the unit of GeV.  The four
  pairs of the extra matters exist at $4$ TeV. }
\label{fig:nf4}
\end{center}
\end{figure}

We also show that contours of the constant Higgs mass for
$N_5=4$ and $M_{N_5}=4$ TeV in fig.~{\ref{fig:nf4}}. The Higgs mass of
about 125 GeV (or larger) is realized in a wide region where the
gluino mass is larger than about $1.2$ TeV.  Correspondingly, the Bino
and Wino masses are larger than 380 GeV and 630 GeV,
respectively. Notice that the squark mass is larger than 3 TeV, which
is too heavy to be observed at the LHC. The sleptons are also heavier
than about 1 TeV. In the calculation, we adapt the universal gaugino
masses at the GUT scale and the Bino is the next-to-the lightest
SUSY particle (NLSP) in the whole region. (For the case without the
GUT relation, see the discussion below.)

\begin{figure}[t]
\begin{center}
\includegraphics[width=7.9cm]{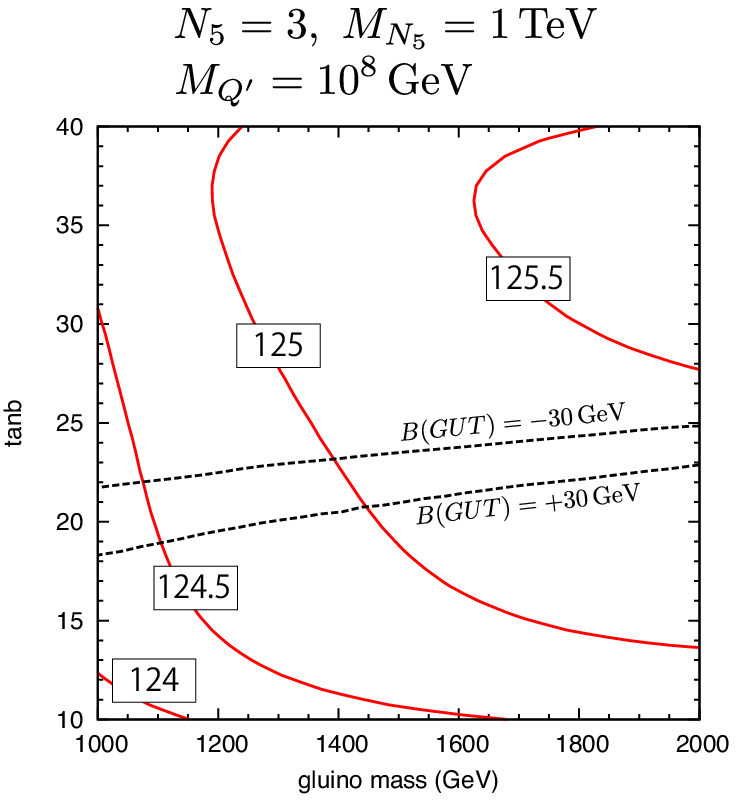}
\includegraphics[width=7.9cm]{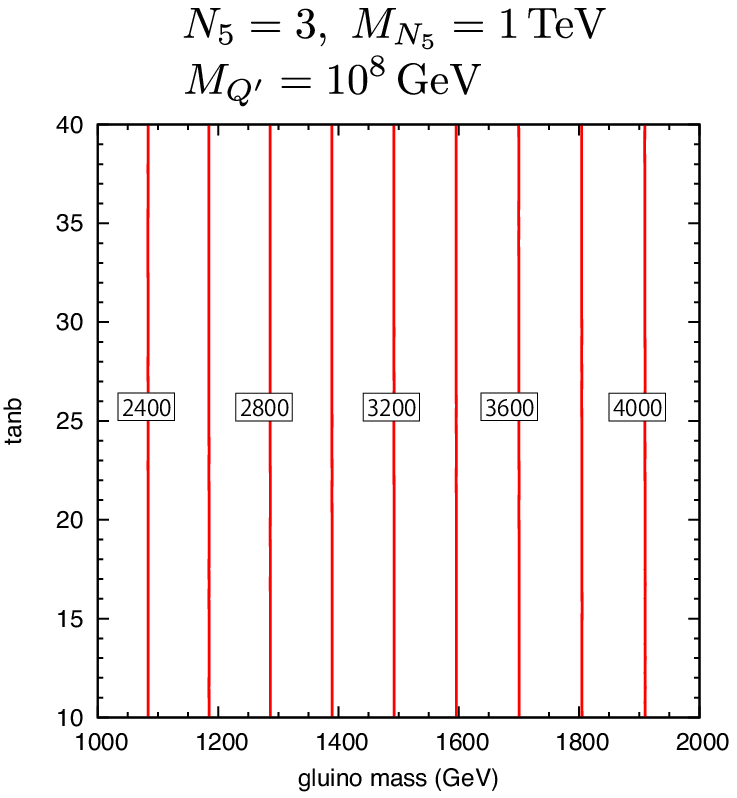}
\caption{Contours of the Higgs mass (left) and the squark mass (right)
  on $m_{\tilde{g}} - \tan\beta$ plane in the unit of GeV.  The three
  pairs of the extra matters exist at $1$ TeV and another pair of the
  extra matters exist at $10^8$ GeV.}
\label{fig:PQ}
\end{center}
\end{figure}

So far, we have seen that the existence of extra matters at the SUSY
scale enhances the Higgs mass in the gauge mediated SUSY breaking
scenario.  More enhancement may be realized if there exist additional
extra matters at the intermediate scale.  One of the motivations to
consider such extra matters at the intermediate scale is SUSY axion
model.  As an example of the enhancement due to the extra matters at
the intermediate scale, we consider the case where there exists one
pair of ${\bf 5}+\bar{\bf 5}$ extra matters at $10^8$
GeV~\footnote{The mass can be smaller than the Peccei-Quinn (PQ)
  symmetry breaking scale, i.e., for instance, Yukawa coupling of
  $\mathcal{O}(10^{-2})$\,$\times$\,PQ breaking scale.} (as well as
three pairs of ${\bf 5}+\bar{\bf 5}$ extra matters at TeV scale).  In
fig.~\ref{fig:PQ}, contours of the Higgs mass (left panel) and squark
mass (right panel) are shown.


%
%

Finally, we comment on cosmological implication of Bino-like
neutralino as the NLSP.  In the early universe, neutralinos are
produced and may decay after the BBN epoch in particular when R-parity
is conserved.  Hadro- and photo-dissociation processes are caused by
the decay of the neutralino, and the success of the BBN may be spoiled
if the lifetime of the neutralino is longer than $\sim 1$ sec
\cite{KKM}.  In the present case where $M_a/m_{3/2}\sim 100$, the
parameter region which we are interested in mostly conflicts with the
BBN constraints.\footnote
{If stau is the NLSP, which may be the case in the gaugino mediation
  model without extra matters, the constraint is much weaker
  \cite{KKM}.  Assuming that stau decays into tau and
  gravitino via supercurrent interaction, the stau mass is required to
  be larger than 200 GeV for $m_{\tilde{\tau}}/m_{3/2}=100$ (with
  $m_{\tilde{\tau}}$ being the mass of stau), in order for successful
  BBN scenario.  In addition, for $m_{\tilde{\tau}}/m_{3/2}=300$, no
  constraint is obtained.  }
Such a problem may be solved if the NLSP is Wino-like neutralino
instead of Bino-like neutralino.\footnote
{The GUT relation among gaugino masses may not hold even if the SM gauge
  couplings are unified.  For example, in the product-group unification
  scenario \cite{productGUT}, this is the case
  \cite{ArkaniHamed:1996jq}.}
In such a case, the thermal relic abundance of the NLSP is
significantly suppressed, and the BBN constraints are relaxed.
Interestingly, if the Wino-like neutralino is the NLSP, the signal of
the Wino production may be observed at the LHC \cite{WinoLHC, Pure2}.
Another possibility to avoid the confliction with the BBN constraints
is to introduce small R-parity violation.

\section{Conclusion and discussion}
In this letter, we have considered a gaugino mediated SUSY breaking
scenario without the Polonyi problem. The gaugino mediation naturally
occurs with the requirements for avoiding the serious Polonyi
problem. With this setup, we have evaluated the Higgs boson mass and
found that the Higgs mass of around 125 GeV may be realized with
gluino mass heavier than about 4 TeV in the MSSM. Unfortunately, such
heavy colored SUSY particles are out of reach of the LHC experiment.

However, if there exist a number of extra matters at $1-10$ TeV, the
Higgs boson mass can be significantly enhanced with relatively small
gluino mass of $1-2$ TeV. This is because the trilinear coupling of
the scalar top is enhanced by the change of the RGEs of the gauge
coupling constants and gaugino masses. With the uncertainty in the
calculation of the Higgs boson mass ($2-3$ GeV), the squark mass of
around 2 TeV becomes consistent with observed value of the Higgs boson
mass. Such gluino and squarks are within the reach of the LHC
experiment.  In addition, it is notable that the masses of other
SUSY particles can be much below $1\ {\rm TeV}$; in the parameter
region of our interest, the masses of slepton, Bino, and Wino are
about $470-670$ GeV, $210-310$ GeV, and $330-490$ GeV, respectively.
These non-colored SUSY particles can be targets of future $e^+e^-$
linear collider experiments.

In this letter, we have concentrated on the scenario in which only the
gauge multiplets (and the inflaton) have enhanced couplings to the
Polonyi field.  From the point of view of solving the SUSY FCNC
problem, the Higgs fields may also strongly couple to the Polonyi
field.  If so, the behaviors of the SUSY breaking parameters may be
significantly altered~\cite{Brummer:2012ns}.  In particular, the cut-off-scale values of the
$A$-parameters and soft SUSY breaking Higgs masses are expected to
become much larger than the gravitino mass.  Such a scenario will be
studied in a separate publication \cite{MorYanYok}.

Our scenario is consistent with the cosmological observation. The
baryon asymmetry can be from non-thermal leptogenesis with relatively
high reheating temperature as $10^6$ GeV. The gravitino is the LSP and
the candidate for dark matter. If the R-parity is conserved, the
gravitino mass of $m_{3/2}\sim 10\ {\rm GeV}$, which is expected in
the present scenario, may conflict with the BBN constraints in
particular when the Bino-like neutralino is the NLSP.  However, the
BBN constraints can be avoided if the NLSP is Wino-like neutralino or
if a small violation of R-parity is introduced. Even with a small
R-parity violation, the gravitino can be long-lived enough to be a
viable candidate for dark matter \cite{Buchmuller:2007ui,
  Takayama:2000uz} .

\section*{Acknowledgment}
This work is supported by the Grant-in-Aid for Scientific research
from the Ministry of Education, Science, Sports, and Culture (MEXT),
Japan, No.\ 22244021 (T.M. and T.T.Y.)  No.\ 23104008 (T.M.), No.\
60322997 (T.M.), and also by the World Premier International Research
Center Initiative (WPI Initiative), MEXT, Japan. The work of N.Y. is
supported in part by JSPS Research Fellowships for Young Scientists.


\end{document}